\documentclass[aps,prl,reprint,superscriptaddress]{revtex4-2}

\usepackage{graphicx}
\usepackage{dcolumn}
\usepackage{bm,braket,here,comment}

\begin{document}

\preprint{}

\title{Observation of the Sign Reversal of the Magnetic Correlation in a Driven-Dissipative Fermi Gas in Double Wells}


\author{Kantaro Honda}
 \email[]{honda.kantaro.35m@st.kyoto-u.ac.jp}
 \affiliation{Department of Physics, Graduate School of Science, Kyoto University, Kyoto 606-8502, Japan}

\author{Shintaro Taie}
  \affiliation{Department of Physics, Graduate School of Science, Kyoto University, Kyoto 606-8502, Japan}

\author{Yosuke Takasu}
  \affiliation{Department of Physics, Graduate School of Science, Kyoto University, Kyoto 606-8502, Japan}

\author{Naoki Nishizawa}
  \affiliation{Department of Physics, Graduate School of Science, Kyoto University, Kyoto 606-8502, Japan}

\author{Masaya Nakagawa}
  \affiliation{Department of Physics, University of Tokyo, 7-3-1 Hongo, Bunkyo-ku, Tokyo 113-0033, Japan}

\author{Yoshiro Takahashi}
  \affiliation{Department of Physics, Graduate School of Science, Kyoto University, Kyoto 606-8502, Japan}

\date{\today}

\begin{abstract}
We report the observation of the sign reversal of the magnetic correlation from antiferromagnetic to ferromagnetic in a dissipative Fermi gas in double wells, utilizing the dissipation caused by on-site two-body losses in a controlled manner.
We systematically measure dynamics of the nearest-neighbor spin correlation in an isolated double-well optical lattice, as well as a crossover from an isolated double-well lattice to a one-dimensional uniform lattice.
In a wide range of lattice configurations over an isolated double-well lattice, we observe a ferromagnetic spin correlation, which is consistent with a Dicke type of correlation expected in the long-time limit.
This work demonstrates the control of quantum magnetism in open quantum systems with dissipation.

\end{abstract}


\maketitle

\begin{figure}[b]
  \includegraphics[width=8.54cm]{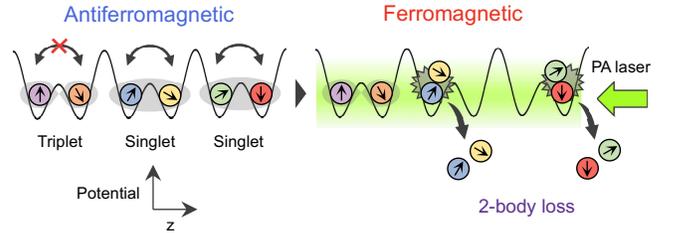}
  \caption{\label{Fig.1} Schematic illustration of the mechanism of the reversal of the magnetic correlation from the antiferromagnetic (left) to the ferromagnetic correlation (right) in dissipative double wells with on-site two-body losses.
Here, we show the case of six spin-components (magnetic quantum number $m_F=-5/2, -3/2, ..., +5/2$) in the system, which is the same situation as in this study.
While double occupancy on the same site occurs in the singlet state, which has a spatially symmetric wave function, it is prohibited by the Pauli exclusion principle in the triplet state, which has a spatially antisymmetric wave function.
Thus, only singlet states are lost under on-site two-body losses, which makes the nearest-neighbor magnetic correlation reverse.
In this work, we induce on-site two-body losses with a PA process.}
\end{figure}
\begin{figure}
  \includegraphics[width=8.54cm]{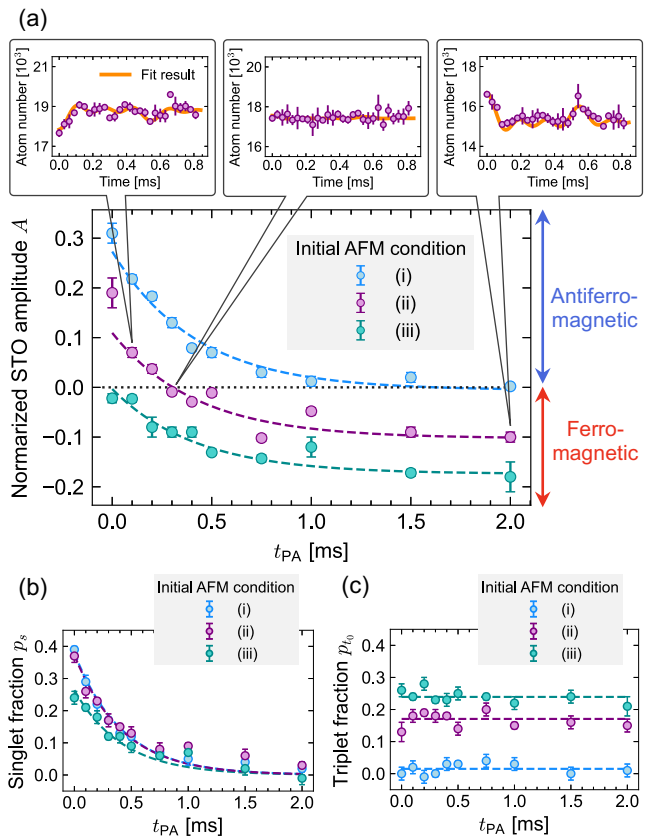}
  \caption{\label{Fig.2} Dynamics of (a) the normalized STO amplitude $A$, (b) the singlet fraction $p_s$, and (c) the triplet fraction $p_{t_0}$ in an isolated double-well lattice with on-site two-body losses.
Here, we show their dependency on the irradiation time $t_{\text{PA}}$ of the PA laser.
In (a), three different initial antiferromagnetic (AFM) conditions are prepared with different atom-loading schemes into an optical lattice (Sec.~S.4 in SM~\cite{SM}).
The dashed curves in (a), (b), and (c) show the results of the exponential fit of the data with an offset, without an offset, and the constant fit, respectively, where the damping time constant of the exponential function is determined by the co-fit to $A$ and $p_s$.
Error bars denote the standard deviation, which is calculated with fitting errors of STO measurement and the standard deviations of the total atom number.
In markup balloons in (a) (upper), measured STO signals for the initial condition (ii) are shown, which correspond to $t_{\text{PA}}=0.1$~ms, $0.3$~ms, and $2.0$~ms in order from the left.
Error bars in STO signals denote the standard deviation of three scans.}
\end{figure}
\begin{figure}
  \includegraphics[width=8.54cm]{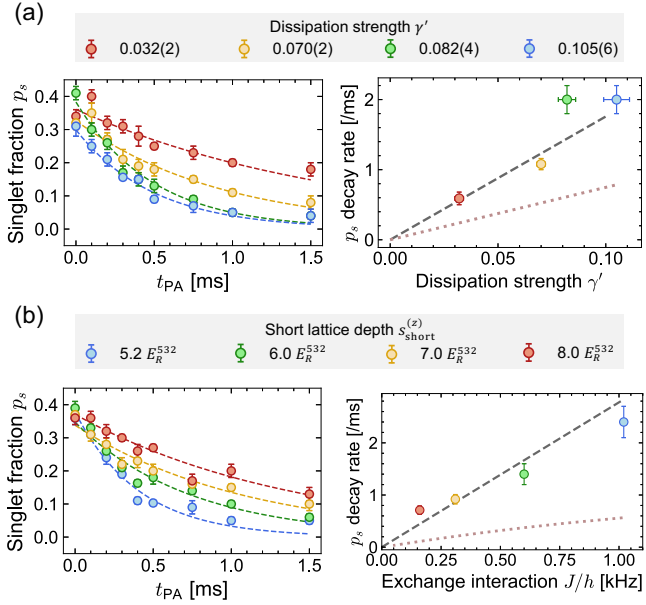}
  \caption{\label{Fig.3} (a) (left) Dynamics of the singlet fraction $p_s$ for each dissipation strength $\gamma^{\prime}$ in a dissipative isolated double-well lattice.
Each dashed curve shows the results of the exponential fit without an offset.
(right) Dependency of the decay rate of $p_s$ obtained by the exponential fit on the dissipation strength $\gamma^{\prime}$.
The gray dashed line shows the result of the linear fit through the origin.
The dotted line shows the result of the calculation considering the finite value of $U/t_d=4$ in our case.
Vertical error bars denote the fitting error of the exponential fit to $p_s$.
Horizontal error bars denote the errors of the measured two-body loss rates.
(b) (left) Dynamics of the singlet fraction $p_s$ for each short lattice depth in a dissipative isolated double-well lattice.
Here, $E_R^{532}$ denotes the recoil energy of the short lattice.
Each dashed line shows the results of the exponential fit without an offset.
(right) Dependency of the decay rate of $p_s$ obtained by the exponential fit on the exchange interaction $J$.
The gray dashed line shows the result of the linear fit through the origin.
The dotted line shows the result of the calculation considering the finite value of $U/t_d$.}
\end{figure}
\begin{figure}
  \includegraphics[width=8.54cm]{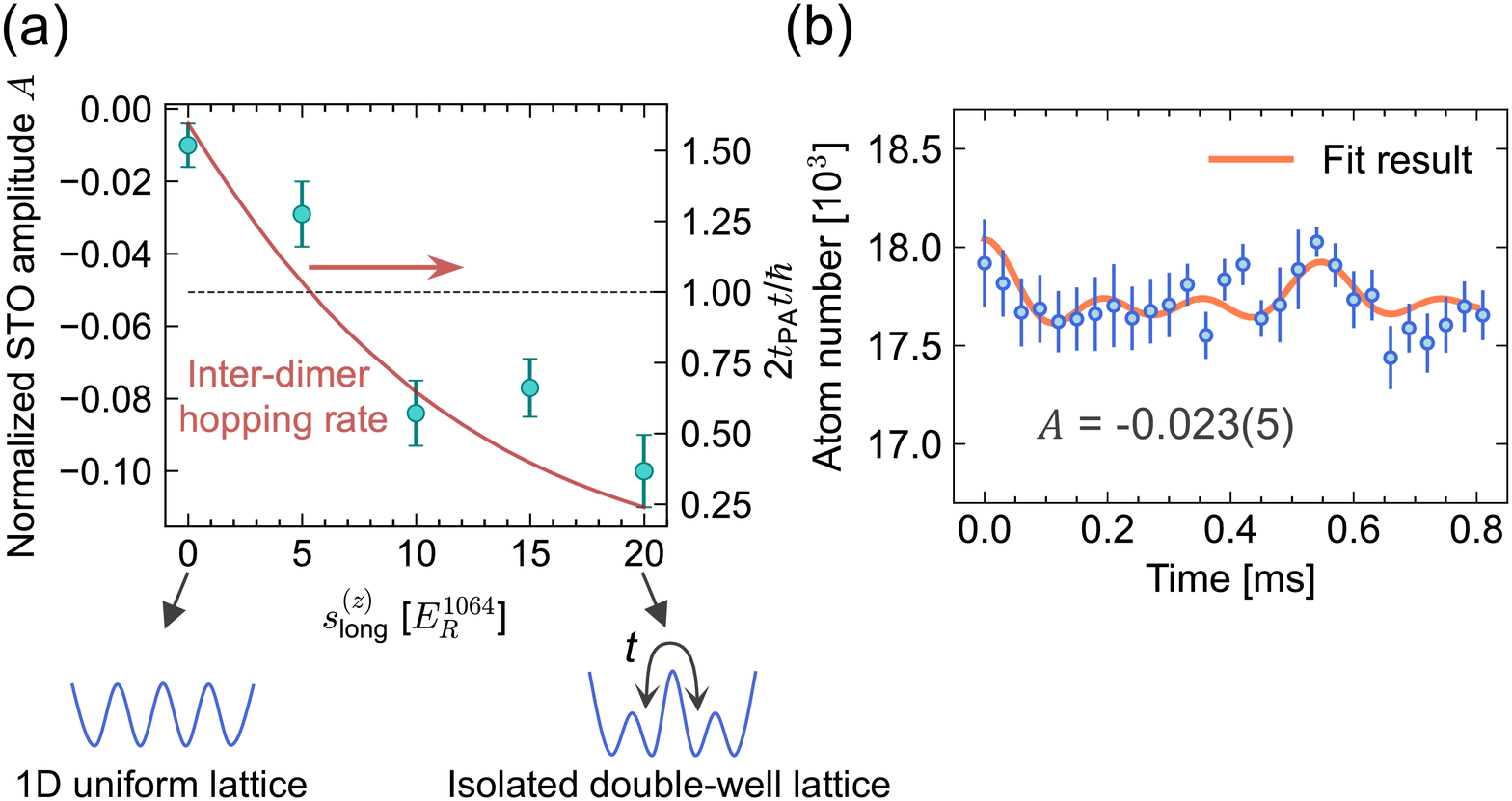}
  \caption{\label{Fig.4} (a) Normalized STO amplitude $A$ in a crossover from an isolated double-well lattice of $s_L=[100, 80, (20.8, 20)]$ to a 1D uniform lattice of $s_L=[100, 80, (20.8, 0)]$.
Here, the dependency of $A$ on the long lattice depth is shown.
We note that the number of scans in STO measurement at $s_{\text{long}}^{(z)}=0 E_R^{1064}$ is seventeen, otherwise three.
The red solid curve shows the quantity $2t_{\text{PA}}t/\hbar$ which is a measure of the number of inter-dimer hopping events during the PA laser irradiation time, and the black dashed line represents the value of 1.
Here, the factor 2 corresponds to tunnelings to double wells on both sides.
(b) Measured STO signal in a 1D uniform lattice of $s_L=[100, 80, (8, 0)]$.
The orange solid curve shows the fit result.
Error bars denote the standard error of twelve scans.}
\end{figure}

Dissipation can play a novel role in preparing and manipulating quantum states of interest \cite{doi:10.1080/00018732.2014.933502, MULLER20121}.
In cold atom systems, which are often regarded as ideal isolated quantum systems, it is also possible to study open quantum systems by making use of their high controllability.
So far, dissipative systems with one- \cite{PhysRevLett.110.035302,PhysRevLett.115.050601,PhysRevLett.116.235302,doi:10.1126/sciadv.aat6539,PhysRevLett.115.140402,PhysRevX.7.011034}, two- \cite{PhysRevLett.95.190406,doi:10.1126/science.1155309,Yan2013,doi:10.1126/sciadv.1701513,PhysRevA.99.031601}, and three-body particle losses \cite{PhysRevLett.92.190401,PhysRevLett.108.215302,PhysRevResearch.2.043050} have been investigated, and part of novel phenomena brought about only by the presence of dissipation has been clarified.\\
\indent
In quantum magnetism, dissipation can also have a great influence on its equilibrium state.
In the previous work, sign reversal of magnetic correlations from antiferromagnetic to ferromagnetic in a dissipative system with on-site two-body losses is theoretically predicted \cite{PhysRevLett.124.147203} (Fig.~\ref{Fig.1}).
For a dissipative system, the formation of a highly entangled Dicke state~\cite{PhysRev.93.99}, which has a fully symmetric spin wave function, is also predicted \cite{PhysRevLett.109.230501,PhysRevA.105.L051302}.
Experimentally, atom loss dynamics in the one-dimensional (1D) dissipative Fermi-Hubbard system has been investigated, and the dynamical generation of the Dicke state has been inferred based on the suppressed atom losses after initial fast transient dynamics \cite{Sponselee_2018}.
However, in Ref.~\cite{Sponselee_2018}, spin correlations, which are a more direct signature of the ferromagnetism characterizing the Dicke state, have not been measured.
Moreover, a fixed strength of dissipation was introduced, and therefore systematic study of dissipative dynamics with controllable strength of dissipation was not carried out.\\
\indent
In this Letter, we implement a driven-dissipative Fermi-Hubbard system in a dimerized lattice with on-site two-body losses.
Engineered dissipation is introduced in a controlled manner by a photoassociation (PA) process~\cite{RevModPhys.78.483} (Fig.~\ref{Fig.1}).
This system is described by the dissipative Fermi-Hubbard model in a dimerized lattice, and the dynamics of its density matrix $\hat{\rho}$ is described by the following quantum master equation~\cite{PhysRevLett.124.147203,PhysRevA.105.L051302}:
\begin{equation}
  \frac{d\hat{\rho}}{d\tau}=\frac{i}{\hbar} [\hat{\rho},\hat{H}] +\frac{1}{2}\sum_{j,\sigma\neq\sigma'}\left(\hat{L}_{j\sigma\sigma'}\hat{\rho} \hat{L}_{j\sigma\sigma'}^{\dagger}-\frac{1}{2}\{ \hat{L}_{j\sigma\sigma'}^{\dagger}\hat{L}_{j\sigma\sigma'},\hat{\rho}\} \right),
\end{equation}
\begin{eqnarray}
  \hat{H}&=&-t_d\sum_{\langle i,j\rangle^{z}_{\textbf{-}},\sigma}
  \left( \hat{c}_{i\sigma}^{\dagger} \hat{c}_{j\sigma}+{\rm H.c.} \right) +\frac{U}{2}\sum_{j,\sigma\neq\sigma'} \hat{n}_{j\sigma}\hat{n}_{j\sigma'}-\mu\sum_{j,\sigma}\hat{n}_{j\sigma} \nonumber \\
  &&-t\sum_{\langle i,j\rangle^{z}_{-},\sigma}
  \left( \hat{c}_{i\sigma}^{\dagger} \hat{c}_{j\sigma}+{\rm H.c.} \right)-t_{xy}\sum_{\langle i,j\rangle^{xy}_{-},\sigma}\left( \hat{c}_{i\sigma}^{\dagger} \hat{c}_{j\sigma}+{\rm H.c.} \right),
\end{eqnarray}
where $\hat{c}_{i\sigma}$ is the fermionic annihilation operator for a site $i$ and spin $\sigma$, $\hat{n}_{j\sigma}=\hat{c}_{j\sigma}^{\dagger}\hat{c}_{j\sigma}$ is the number operator, $t_d$, $t$, and $t_{xy}$ are the tunneling amplitudes between nearest-neighbor sites in the intra-dimer bond $\langle i,j\rangle^{z}_{\textbf{-}}$, the inter-dimer bond $\langle i,j\rangle^{z}_{-}$ along the $z$-axis, and the inter-dimer bond $\langle i,j\rangle^{x(y)}_{-}$ along the $x(y)$-axis, respectively, $U$ is the on-site interaction strength, $\mu$ is the chemical potential, and $\hat{L}_{j\sigma\sigma'}=\sqrt{2\gamma}\hat{c}_{j\sigma}\hat{c}_{j\sigma'}$ is the Lindblad operator, which describes the on-site two-body loss with the loss rate $\gamma$.
In dissipative isolated double-well systems, we systematically measure the dynamics of nearest-neighbor spin correlations.
Moreover, we measure spin correlations in a crossover from an isolated double-well lattice to a 1D uniform lattice. We observe a sign-reversed magnetic correlation, which is consistent with a Dicke type of correlation expected in the long-time limit.\\
\indent
{\it Experimental setup.{\bf \textemdash}}
A degenerate Fermi gas of six-component $^{173}$Yb ($T/T_F\sim 0.15$, $T_F$: Fermi temperature), which is generated by evaporative cooling, is loaded into an optical superlattice which realizes a dimerized or a 1D optical lattice as well as the crossover between them.
Our optical lattice potential is given by
\begin{eqnarray}
  V(x,y,z)&=&-V_{\text{short}}^{(z)}\cos^2(2k_Lz+\pi/2)-V_{\text{long}}^{(z)}\cos^2(k_Lz) \nonumber \\
  &&-V_{\text{short}}^{(x)}\cos^2(2k_Lx)-V_{\text{short}}^{(y)}\cos^2(2k_Ly),
\end{eqnarray}
where $k_L=2\pi/\lambda$ with $\lambda=1064~\text{nm}$ is the wave number of the long lattice ($V_{\text{long}}^{(z)}=0$ for a 1D uniform optical lattice).
In the following, we represent the potential depth of the optical lattice as $s_L=\left[s_{\text{short}}^{(x)}, \ s_{\text{short}}^{(y)},\ (s_{\text{short}}^{(z)},\ s_{\text{long}}^{(z)})\right]=\left[V_{\text{short}}^{(x)}, \ V_{\text{short}}^{(y)},\ (V_{\text{short}}^{(z)},\ V_{\text{long}}^{(z)})\right]/E_R$, where $E_R=\hbar^2k_L^2/2m$ is the recoil energy for the long lattice.\\
\indent
In this work, we introduce on-site two-body losses as dissipation utilizing a PA technique.
In the PA process, two ground-state atoms in a doubly occupied site where a PA laser beam is irradiated are converted to an electronically excited short-lived molecule, which rapidly escapes from an optical trap~\cite{Taie2012,doi:10.1126/sciadv.1701513,PhysRevLett.121.225303,Taie2022}.
In this way, on-site two-body losses are realized with the PA laser, enabling us to control the dissipation strength by controlling the PA laser intensity.
Use of the deep PA resonance, far detuned by -5.76~GHz from the ${^1}S_0\leftrightarrow{^3}P_1\ (F'=7/2)$ transition, is effective to suppress the one-body loss and heating due to photon scattering.\\
\indent
To directly measure nearest-neighbor spin correlations, we utilize a singlet-triplet oscillation (STO)~\cite{PhysRevLett.105.265303} which is optically induced with a spin-dependent potential gradient [see Sec.~S.1 in the Supplemental Material (SM) \cite{[{See Supplemental Material at [URL will be inserted by publisher] for the experimental details and the calculation method which includes Refs.~\cite{[{[First reference in Supplemental Material not already in paper] }]PhysRevA.90.013614,PhysRevLett.121.225303,Taie2022,doi:10.1126/sciadv.1701513,[{[Fifth reference in Supplemental Material not already in paper] }]PhysRevA.66.061403,[{[Sixth reference in Supplemental Material not already in paper] }]PhysRevA.71.013417,[{[Seventh reference in Supplemental Material not already in paper] }]Haimberger_2006,[{[Eighth reference in Supplemental Material not already in paper] }]PhysRevLett.101.060406,PhysRev.103.20,[{[Tenth reference in Supplemental Material not already in paper] }]doi:10.1126/science.1227831,doi:10.1080/00018732.2014.933502,PhysRevLett.124.147203}}]SM} for the details of the STO measurement].
A typical STO signal for a six-component mixture in a dimerized lattice is shown in Fig.~\ref{Fig.2}(a) (upper left).
To quantitatively evaluate the magnetic correlation, we fit the data with a function
\begin{equation}
  f(t)=-\frac{a}{3}e^{-t/\tau}\left[\cos(\omega t)+\cos(2\omega t)+\cos(3\omega t)\right]+b,\label{fitfunc}
\end{equation}
where $a, b, \tau, \omega$ are fitting parameters.
Here, the functional form with three frequency components with a simple integer ratio is related with the linear polarization of the beam which induces a spin-dependent potential gradient.
More specifically, the linearly polarized gradient beam makes the differential light shift the same for the spin pairs of ($|m_F|$ = 1/2, 3/2) with the corresponding STO frequency $\omega_1$, (3/2, 5/2) with $\omega_2$, and (1/2, 5/2) with $\omega_3$.
Note that, in this configuration, three spin pairs with the same $|m_F|$, namely, ($m_F$=-1/2, 1/2), (-3/2, 3/2), and (-5/2, 5/2), do not show the STO because the differential light shifts due to the gradient beam are zero for these spin pairs.
In addition to this, a simple relation $\omega_1:\omega_2:\omega_3=1:2:3$ holds between these frequencies due to the properties of the Clebsch-Gordan coefficients for the associated transition.
Along with the STO data, we separately measure the total atom number $N_{\text{tot}}$ in the lattice with no PA laser for dissipation.
Similarly, we also measure the total atom number $N$ in the lattice after the irradiation of the PA laser for dissipation.
We quantify the magnetic correlation by the normalized STO amplitude $A$, and the number of singlet pairs $N_s$ and triplet pairs $N_{t_0}$, and their fractions $p_s=N_s/N_{\text{tot}}$ and $p_{t_0}=N_{t_0}/N_{\text{tot}}$:
\begin{eqnarray}
  A=&&(4/5)(N_s-N_{t_0})/N_{\text{tot}}\ =2a/N_{\text{tot}}, \label{STOamp} \\
  p_s=&&N_s/N_{\text{tot}}\ =(N+a-b)/N_{\text{tot}}, \label{STOps} \\
  p_{t_0}=&&N_{t_0}/N_{\text{tot}}=(2N-3a-2b)/2N_{\text{tot}}. \label{STOpt0}
\end{eqnarray}
Here, the factor 4/5 in Eq.~(\ref{STOamp}) reflects the fraction of the number of spin pairs contributing to the STO (see Sec.~S.2 in SM~\cite{SM} for the derivation of Eqs.~(\ref{STOamp})-(\ref{STOpt0})).
We note that, instead of measuring the atom number of each spin component, the number of singlet pairs $N_s$ and triplet pairs $N_{t_0}$ are calculated from Eqs.~(6) and (7) through fitting the data with Eq.~(4) under the assumption of SU(6) symmetry, namely, that all possible spin pairs equally contribute to the spin correlation.
Note that the correspondence between the sign of slope of STO signals at initial time and the spin correlation can be understood by the correspondence between the coefficient $a$ in the fitting function Eq.~(4) and the STO amplitude $A$ in Eq.~(5).\\
\indent
{\it Dynamical reversal of the magnetic correlation in an isolated double-well lattice.{\bf \textemdash}}
To reveal the effects of the on-site two-body loss on the magnetic correlation, we first measure the spin correlation dynamics in a dissipative isolated double-well lattice.
Figures~\ref{Fig.2}(a), \ref{Fig.2}(b), and \ref{Fig.2}(c) show the results of the normalized STO amplitude $A$, the singlet fraction $p_s$, and the triplet fraction $p_{t_0}$, respectively.
The STO data are taken for the atom number $N_{\text{tot}}=2.7\times10^4$, which is small enough to suppress double occupancies.
Dimensionless dissipation strength $\gamma^{\prime}=\hbar\gamma/U=\hbar(\gamma_{\text{PA}}/2)/U$ is 0.082(4), where $\gamma_{\text{PA}}$ is the two-body loss rate due to the PA laser ($\gamma_{\text{PA}}=4.2(2)$~/ms) and is measured from atom loss dynamics in an isolated double-well lattice (Sec.~S.3 in SM~\cite{SM}).
We note that while $\gamma$ and $\gamma_{\text{PA}}$ are tunable via the PA laser power and detuning as well as the lattice depth, $\gamma'$ does not depend on the lattice depth.
In this measurement, the dependency of the spin correlation dynamics on the magnitude of initial ($t_{\text{PA}}=0$) antiferromagnetic correlations is also investigated, which is realized by changing atom-loading schemes into an isolated double-well lattice (Sec.~S.4 in SM~\cite{SM}).
In Fig.~\ref{Fig.2}(a), the normalized STO amplitude decreases as the dissipation is introduced, regardless of the initial conditions.
Especially, in the initial condition (ii), the form of the STO signal reverses due to the dissipation, indicating the sign of $A$ changing from positive to negative (Fig.~\ref{Fig.2}: upper left to right).
This means that the magnetic correlation between the nearest neighbor sites in the isolated double-well lattice changes from an antiferromagnetic to ferromagnetic correlation, which has been theoretically predicted~\cite{PhysRevLett.124.147203} and corresponds to the negative spin temperature state \cite{PhysRev.103.20} (Sec.~S.5 in SM~\cite{SM}).
Note that the photon scattering by the PA laser, which can decrease spin correlations, is negligible on the time scale of this measurement.
Figure~\ref{Fig.2}(b) shows that the singlet fraction $p_s$ decreases by the application of the dissipation, while the triplet fraction $p_{t_0}$ shows no apparent decrease, as shown in Fig.~\ref{Fig.2}(c).
These results are in good agreement with the theoretical consideration that the change of the spin correlation is solely due to the decrease of the number of the singlet state while that of the triplet state is kept constant.
Note that the observed behaviors originating from two-body physics inside the double-well are common, irrespective of SU(2) or SU(6) system.
The dissipative SU(6) system realized in this work opens the door for the study of non-ergodic dynamics in the dissipative SU($\mathcal{N}$) Fermi-Hubbard model \cite{arXiv:2205.07235v1}.\\
\indent
{\it Dependency of the dissipative dynamics on the dissipation strength and the exchange interaction.}{\bf \textemdash}
Next, we investigate the dependency of the spin correlation dynamics in an isolated double-well lattice on the dissipation strength $\gamma^{\prime}$ and the exchange interaction $J=4t_d^2/U$.
When $U\gg t_d$, the loss rate of the singlet state is theoretically given as $2\Gamma/\hbar$ where $\Gamma=J\gamma^{\prime} /(1+{\gamma^{\prime}}^2)$ \cite{PhysRevLett.124.147203}.
Considering $\gamma^{\prime}\ll 1$ for this measurement, we expect the linear dependency of the decay rate of $p_s$ on $\gamma^{\prime}$ and $J$.
We note that the continuous Quantum Zeno effect, which suppresses $\Gamma$ at $\gamma'\gg 1$, does not play a role in this measurement with $\gamma'\ll 1$.
Figure~\ref{Fig.3}(a) (left) shows the result of the singlet fraction $p_s$ measured at each dissipation strength $\gamma^{\prime}$.
One can clearly see that the larger the dissipation strength is, the faster $p_s$ decays.
In Fig.~\ref{Fig.3} (a) (right), we show the dependency of the decay rate of $p_s$ on the dissipation strength $\gamma^{\prime}$.
We find that the decay rate of $p_s$ increases approximately in proportion to $\gamma^{\prime}$.
The dotted line in Fig.~\ref{Fig.3}(a) (right) represents the result of the calculation considering the finite value of $U/t_d=4$ in our case, which also shows the linear dependence, similar to the experimental data (see Sec.~S.6 in SM~\cite{SM} for the details of the calculation).
Figure~\ref{Fig.3}(b) (left) shows the result of the singlet fraction $p_s$ measured at each short lattice depth, where the dissipation strength is fixed at $\gamma^{\prime}=0.082(4)$.
The shallower the short lattice depth is, namely, the larger the magnitude of the exchange interaction $J$ is, the faster $p_s$ decays.
We note that $U/t_d$ changes from 4.0 to 10.9 when $s_{\text{short}}^{(z)}$ changes from 5.2 to 8.0.
In Fig.~\ref{Fig.3}(b) (right), we show the dependency of decay rate of $p_s$ on the exchange interaction $J$.
We find that the decay rate of $p_s$ increases approximately in proportion to $J$, which is qualitatively consistent with the calculated results showing approximate linear dependence of $p_s$ on $J$ (see Sec.~S.6 in SM~\cite{SM} for the dependency of the calculated loss rate of the singlet state on $J$ considering the finite value of $U/t_d$ in our case).\\
\indent
{\it Crossover from an isolated double-well lattice to a 1D uniform lattice.{\bf \textemdash}}
Finally, we measure spin correlations in the presence of the dissipation in a crossover from an isolated double-well lattice to a 1D uniform lattice.
Figure~\ref{Fig.4}(a) shows the result of the dependency of the normalized STO amplitude $A$ on the long lattice depth.
Here we prepare the sample in the initial condition corresponding to (iii) in Fig.~\ref{Fig.2}, and therefore the initial spin correlation is expected to be almost zero or small antiferromagnetic one.
In this measurement, the dissipation strength $\gamma^{\prime}$ is fixed to 0.082(4).
We note that we restrict the irradiation time of the PA laser $t_{\text{PA}}$ to 0.5~ms in this measurement to suppress the effect of hopping between 1D tubes and photon scattering by the PA laser.
As shown in Fig.~\ref{Fig.4}(a), we find that the formation of the ferromagnetic spin correlation is suppressed for the smaller long-lattice depths at which atoms can tunnel to neighboring double wells during the irradiation time $t_{\text{PA}}=0.5$~ms of the PA laser.
The quantity $2t_{\text{PA}}t/\hbar$ is a measure of the number of inter-dimer hopping events during the PA laser irradiation time, which is plotted as the red solid curve in Fig.~\ref{Fig.4}(a), and the black dashed line represents the value of 1.
It can be qualitatively understood that this suppression is due to the combined effect of on-site two-body losses and holes created by tunneling between double wells.
The previous work~\cite{PhysRevLett.124.147203} has also pointed out that numerically calculated spin correlations between nearest-neighbor sites of a 1D uniform lattice are smaller than those of an isolated double-well lattice.
Special emphasis is given to the result for the 1D uniform lattice ($s_{\text{long}}^{(z)}=0$), where $A=-0.010(6)$ is small but definitely nonzero and negative.
In Fig.~\ref{Fig.4}(b), we show the STO signal for another 1D uniform lattice, where $A=-0.023(5)$ is negative as well.
These results clearly indicate that in the 1D uniform lattice, the magnetic correlation also reverses, and the ferromagnetic correlation emerges.
For all lattice configurations dealt with in this measurement, the steady state is expected to be the Dicke state.
However, steady states are considered unachieved in this measurement as the right axis in Fig.~\ref{Fig.4}(a) shows that the irradiation time of the PA laser is comparable with the inter-dimer hopping time, which is the consequence of the restriction of the irradiation time of the PA laser for the reasons mentioned above.
So far, in a dissipative 1D uniform lattice, the formation of the Dicke state, which has a fully symmetric spin wave function and shows a ferromagnetic correlation, has only been indirectly suggested based on the atom loss dynamics \cite{Sponselee_2018}.
The observed ferromagnetic correlation in this work provides a signature of formation of Dicke type of correlations in the long-time limit.\\
\indent
{\it Summary and outlook.{\bf \textemdash}}
We have systematically investigated the effects of dissipation on the quantum magnetism and observed the reversed magnetic correlation in a wide range of lattice configurations.\\
\indent
In higher dimensions (2D and 3D), the spin-charge separation does not occur.
Thus, the behavior of the long-range magnetic correlation in a dissipative 2D or 3D system is an interesting problem as a future outlook.\\
\\
\indent
We acknowledge Masahito Ueda for helpful discussions.
This work was supported by the Grant-in-Aid for Scientific Research of JSPS (Nos. JP17H06138, JP18H05405, JP18H05228, JP20K14383, JP21H01014), the Impulsing Paradigm Change through Disruptive Technologies (ImPACT) program, JST CREST (No. JP-MJCR1673), and MEXT Quantum Leap Flagship Program (MEXT Q-LEAP) Grant No. JPMXS0118069021.

\providecommand{\noopsort}[1]{}\providecommand{\singleletter}[1]{#1}%

\end{document}


\preprint{}


\begin{widetext}
\clearpage
\begin{center}
\textbf{\large Supplemental Material for \\Observation of the Sign Reversal of the Magnetic Correlation in a Driven-Dissipative Fermi Gas in Double Wells}
\end{center}
\end{widetext}
\setcounter{figure}{0}
\setcounter{table}{0}
\setcounter{page}{1}
\renewcommand{\theequation}{S\arabic{equation}}
\renewcommand{\thefigure}{S\arabic{figure}}
\renewcommand{\bibnumfmt}[1]{[S#1]}
\renewcommand{\citenumfont}[1]{S#1}

\section{S.1 Singlet-triplet oscillation measurement}
\begin{figure}
  \includegraphics[width=8.54cm]{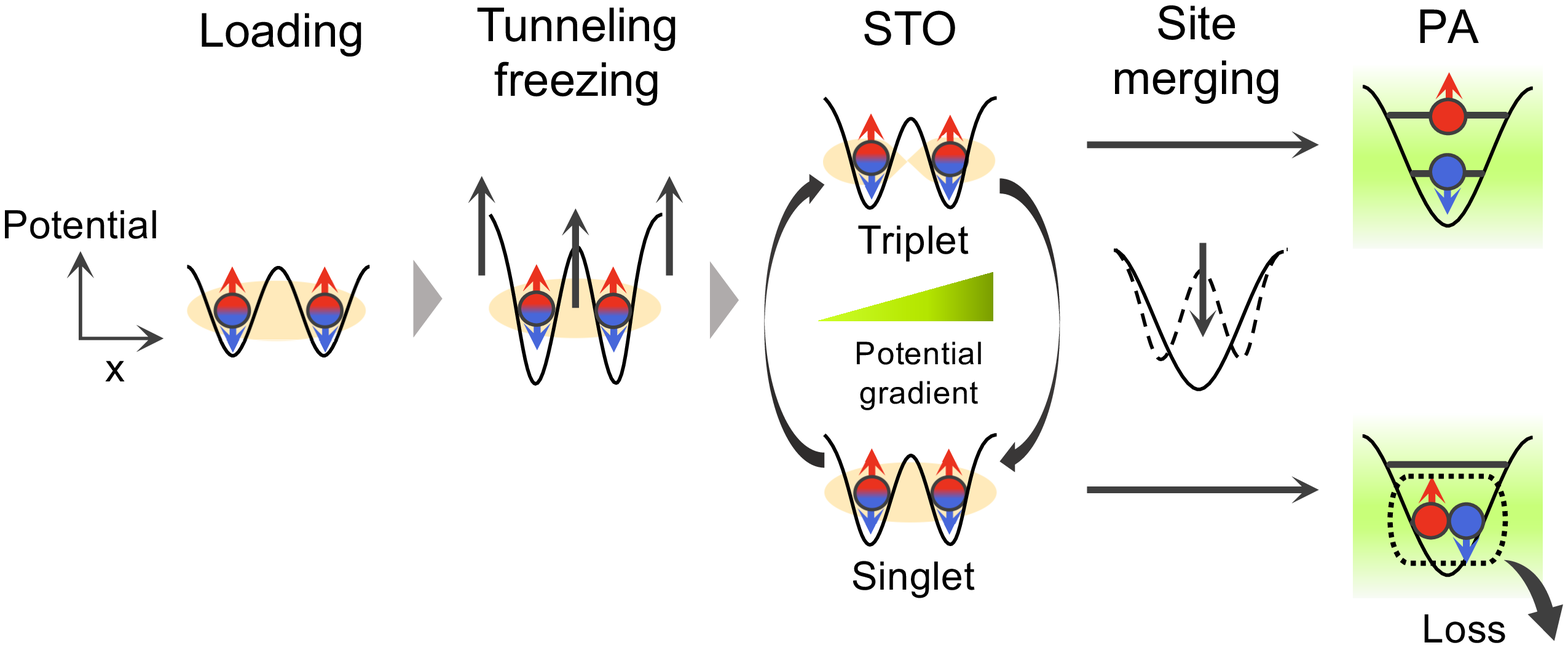}
  \caption{\label{Fig.S1} Schematic illustration of the sequence for STO measurement in a dimer.
Here, we show the case of two spins (red and blue) per dimer.
The sequence flows as loading, tunneling freezing, STO, site merging, and PA.
Depending on the STO time, the two spins form the state in which one spin occupies the lowest band and the other spin occupies the first excited band, or the state in which both spins occupy the lowest band. Atoms in the latter state are removed by the PA laser.}
\end{figure}

The STO measurement is performed after the atom-loading to an optical superlattice in 150~ms and the irradiation of the PA laser to induce dissipation.
Typical parameters for the optical superlattice are $s_L=\left[80, 80, (20.8, 20)\right]$, and the corresponding Hubbard parameters are $t_d/h=1.0$~kHz, $U/t_d=4.0$, $t/h=38$~Hz, and $t_{xy}/h=10$~Hz.
Here, the tunneling amplitudes along the dimerized lattice are determined by fitting a tight-binding model to the energy bands obtained from the first principle calculation for the superlattice potential.
For the on-site interaction strength, we constructed the Wannier function with the method described in Ref.~\cite{PhysRevA.90.013614}.
The typical sequence of the STO measurement is as follows (Fig.~\ref{Fig.S1}).
Note that we call a probed dimer the specific neighboring sites for measuring the spin correlation.\\
\indent
(1) Tunneling freezing : we freeze tunneling between intra- and inter- dimer sites by ramping up the long lattice depth of double wells to $s_L=(80, 80, (20.8, 25))$ in 0.5~ms, followed by ramping up the short lattice depth to $s_L=(80, 80, (80, 25))$ in 10~ms.
We note that the adiabaticity of this ramp-up sequence is confirmed by numerical calculation (Supplemental Material of Ref.~\cite{PhysRevLett.121.225303}).\\
\indent
(2) STO : we drive STOs in probed dimers by irradiating a gradient beam, which induces a spin-dependent potential gradient for atoms.
This potential gradient creates an energy difference $\Delta$ for atoms with different spins in a probed dimer, and the spin states in the dimer coherently oscillate between the singlet state $\ket{s}=(\ket{\sigma_1, \sigma_2}-\ket{\sigma_1, \sigma_2})/\sqrt{2}$ and the triplet state $\ket{t_0}=(\ket{\sigma_1, \sigma_2}+\ket{\sigma_1, \sigma_2})/\sqrt{2}$ at a frequency $\Delta/\hbar$ (STO frequency), where $\sigma_i (i=1,2)$ denotes a spin component.
In this work, we use a linearly polarized light with 2.7~GHz blue detuning from the ${^1}S_0\leftrightarrow{^3}P_1\  (F'=7/2)$ transition as the gradient beam, which makes the differential light shift the same for the spin pairs of $(|m_F|=1/2, 3/2)$ with the corresponding STO frequency $\omega_1$, ($3/2, 5/2$) with $\omega_2$, and ($1/2, 5/2$) with $\omega_3$.
Note that an STO does not occur for the spin pairs with the same $|m_F|$, namely, $(m_F=-1/2, 1/2), (-3/2, 3/2)$, and $(-5/2, 5/2)$ because the differential light shifts are zero for these spin pairs.
In addition to this, a simple relation $\omega_1:\omega_2:\omega_3=1:2:3$ holds between these frequencies due to the properties of the Clebsch-Gordan coefficients for the associated transition.
Note that this relation holds independent of the detuning of the gradient beam.
We also note that actually observed STO signals suffer from the decoherence mainly due to the spatial inhomogeneity of the gradient beam. 
This effect is taken into account in an exponential decay term in the fitting function (Eq.~(4) in the main text).\\
\indent
(3) Site merging : we ramp down the short lattice depth to $s_L=(80, 80, (0, 25))$ in 1~ms.
Here, two atoms in the singlet state which has a spatially symmetric two-particle wave function both occupy the lowest band, while for two atoms in the triplet state which has a spatially antisymmetric two-particle wave function, one atom occupies the lowest band and the other atom occupies the first excited band.\\
\indent
(4) Irradiation of the PA laser : we irradiate the PA laser and remove two atoms which doubly occupy the lowest band, corresponding to the detection of the singlet states.
The PA laser in the STO sequence is detuned by -812~MHz from the ${^1}S_0\leftrightarrow{^3}P_1\ (F'=7/2)$ transition~\cite{PhysRevLett.121.225303,Taie2022}.
Note that we confirm that the PA line for the STO is active for all spin pairs.
After irradiating the PA laser, the absorption imaging follows, in which observed atom number dynamics reflects the magnetic correlation of the prepared sample.

\section{S.2 Derivation of the expressions of STO amplitude $A$ and singlet/triplet fraction $p_s/p_{t_0}$}
In this section, we derive the expressions of STO amplitude $A$ and singlet (triplet) fraction $p_s$ ($p_{t_0}$) (Eqs.~(5)-(7) in the main text).
First, we derive an expression of the STO signal for the case of two spins $(\sigma_1,\sigma_2)$ in a dimer described by the Heisenberg model.
If a spin-dependent potential at a dimer induces an energy difference $\Delta=\hbar \omega$ ($\omega$:\ STO frequency) between the states $\ket{\sigma_1,\sigma_2}$ and $\ket{\sigma_2,\sigma_1}$, the time evolution of the state initially in the singlet state $\ket{s}=(\ket{\sigma_1, \sigma_2}-\ket{\sigma_1, \sigma_2})/\sqrt{2}$ and that of the state initially in the triplet state $\ket{t_0}=(\ket{\sigma_1, \sigma_2}+\ket{\sigma_1, \sigma_2})/\sqrt{2}$ are as follows:
\begin{align}
  \begin{pmatrix}
  \ket{s(t)} \\
  \\
  \ket{t_0(t)} \\
  \end{pmatrix}
  &=\frac{1}{\sqrt{2}}
  \begin{pmatrix}
  \ket{\sigma_1,\sigma_2}- e^{-i\omega t}\ket{\sigma_2,\sigma_1}\\
  \\
  \ket{\sigma_1,\sigma_2}+ e^{-i\omega t}\ket{\sigma_2,\sigma_1}\\
  \end{pmatrix}\\
  &=
  \begin{pmatrix}
  \cos\cfrac{\omega t}{2}\ket{s}-i\sin\cfrac{\omega t}{2}\ket{t_0} \\
  \\
  \sin\cfrac{\omega t}{2}\ket{s}+i\cos\cfrac{\omega t}{2}\ket{t_0} \\
  \label{STOdev1}
  \end{pmatrix}.
\end{align}
Here, if the probabilities of being in the singlet state and the triplet state at $t=0$ are $P_s, P_{t_0}$ respectively, the probability of observing the singlet state after the time evolution by time $t$ is given from Eq.~(\ref{STOdev1}) by
\begin{align}
  P_s(t)&={\rm Tr}[ \rho(t) \ket{s}\bra{s}] \notag \\
  &=P_s\left| \braket{s|s(t)} \right|^2+P_{t_0}\left| \braket{s|t_0(t)} \right|^2 \notag \\
  &=\cfrac{P_s+P_{t_0}}{2}+\cfrac{P_s-P_{t_0}}{2}\cos(\omega t),\label{STOdev2}
\end{align}
where $\rho(t)=P_s\ket{s(t)}\bra{s(t)}+P_{t_0}\ket{t_0(t)}\bra{t_0(t)}$ is the density operator at time $t$.
In the STO measurement sequence, we first drive STOs in probed dimers, remove singlet states selectively with the PA laser, and perform the absorption imaging (see Sec.~S.1).
Thus, the atom number dynamics $N_{\text{STO}}(t)$ (STO signal) observed after the STO measurement sequence is given from Eq.~(\ref{STOdev2}) by
\begin{align}
  N_{\text{STO}}(t)&=N-N_s(t)=N\left[1-P_s(t)\right] \notag \\
  &=\left( N-\cfrac{N_s+N_{t_0}}{2}\right)-\cfrac{N_s-N_{t_0}}{2}\cos(\omega t),
\end{align}
where $N$ is the total atom number and $N_s=NP_s\ (N_{t_0}=NP_{t_0})$ is the number of atoms being initially the singlet (triplet) state.

This expression of the STO signal for the case of two spin components is generalized to the case of $\mathcal{N}(>2)$ spin components.
In this case, $_{\mathcal{N}}{\text{C}}_2$ (number of possible spin pairs) different STO frequencies generally appear in the STO signal and the form of the STO signal is given by
\begin{align}
  N_{\text{STO}}(t)=&N-\sum_{(\alpha,\beta)}\left[\cfrac{N_s^{\alpha\beta}+N_{t_0}^{\alpha\beta}}{2}\right. \notag \\
  &+\left.\cfrac{N_s^{\alpha\beta}-N_{t_0}^{\alpha\beta}}{2}\cos(\omega_{\alpha\beta} t)\right], \label{STOdev3}
\end{align}
where $N_s^{\alpha\beta}, N_{t_0}^{\alpha\beta}$ are the atom numbers being initially the singlet state and the triplet state of spin pair $(\alpha, \beta)$ respectively, and $\omega_{\alpha\beta}$ is the STO frequency for the spin pair $(\alpha, \beta)$.
In our STO measurement sequence, $\mathcal{N}=6$ and only three STO frequencies $\omega_1, \omega_2, \omega_3$ (each for four spin pairs) with a simple relation $\omega_1:\omega_2:\omega_3=1:2:3$ appear (see Sec.~S.1).
With this fact and the assumption of SU($\mathcal{N}$) symmetry in the spin correlation, namely, that all possible spin combinations equally contribute to the spin correlation (i.e., $N_s^{\alpha\beta}=N_s/_{\mathcal{N}}{\text{C}}_2$ and $N_{t_0}^{\alpha\beta}=N_{t_0}/_{\mathcal{N}}{\text{C}}_2$), the STO signal is given from Eq.~(\ref{STOdev3}) by
\begin{align}
  N_{\text{STO}}(t)&=-c\cfrac{N_s-N_{t_0}}{2}\ \cfrac{1}{3}\left[\cos(\omega t)+\cos(2\omega t)+\cos(3\omega t)\right] \nonumber \\
  &+\left[ N-\cfrac{N_s+N_{t_0}}{2}-(1-c)\cfrac{N_s-N_{t_0}}{2}\right] \label{STOdev4},
\end{align}
where $\omega$ is the smallest STO frequency and $c=12/15=4/5$ is the fraction of spin pairs in which STO occurs out of the possible spin pairs.
This form of the STO signal leads to the fitting function to the experimentally obtained data (Eq.~(4) in the main text):
\begin{equation}
  f(t)=-\frac{a}{3}e^{-t/\tau}\left[\cos(\omega t)+\cos(2\omega t)+\cos(3\omega t)\right]+b. \label{STOdev5}
\end{equation}

To quantify the spin correlation, we use the normalized STO amplitude $A=c(N_s-N_{t_0})/N_{\text{tot}}$, the singlet fraction $p_s=N_s/N_{\text{tot}}$, and the triplet fraction $p_{t_0}=N_{t_0}/N_{\text{tot}}$, which are normalized by the total atom number $N_{\text{tot}}$ in the lattice with no PA laser for dissipation.
By comparing corresponding terms in Eq.~(\ref{STOdev4}) ($N$: the total atom number in the lattice after the irradiation of the PA laser for dissipation) with those in Eq.~(\ref{STOdev5}), we obtain the expressions of $A$, $p_s$, and $p_{t_0}$ with the fitting parameters $a$ and $b$ (the same as Eqs.~(5)-(7) in the main text):
\begin{eqnarray}
  A=&&(4/5)(N_s-N_{t_0})/N_{\text{tot}}\ =2a/N_{\text{tot}}, \\
  p_s=&&N_s/N_{\text{tot}}\ =(N+a-b)/N_{\text{tot}}, \\
  p_{t_0}=&&N_{t_0}/N_{\text{tot}}=(2N-3a-2b)/2N_{\text{tot}}.
\end{eqnarray}

\section{S.3 Loss rate measurement of PA}
\begin{figure}
  \includegraphics[width=8.54cm]{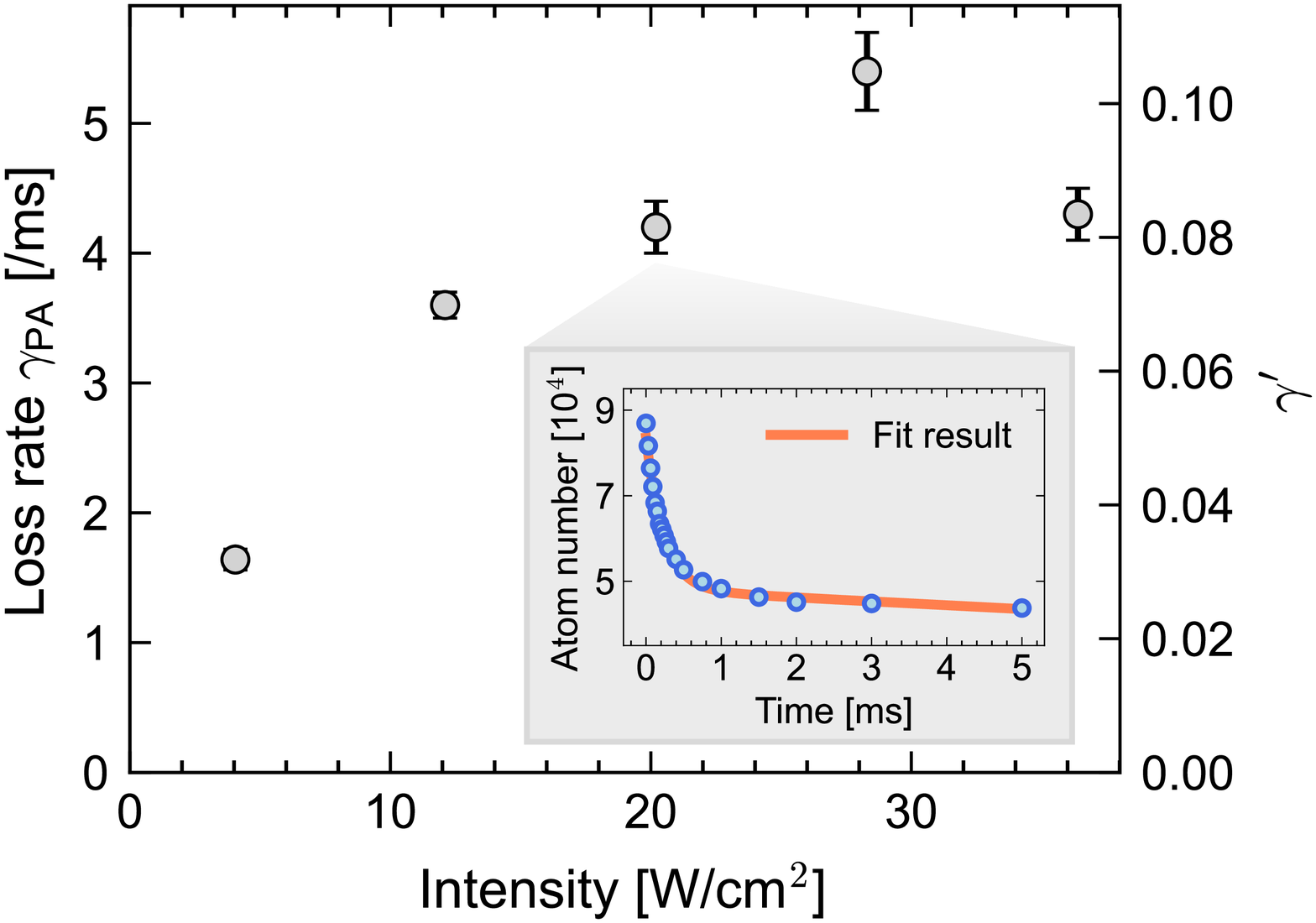}
  \caption{\label{Fig.S2} Two-body loss rate $\gamma_{\text{PA}}$ due to the PA laser.
Here, we show the dependency of the loss rate $\gamma_{\text{PA}}$ (dimensionless dissipation strength $\gamma^{\prime}$ in the right axis) on the intensity of the PA laser.
Error bars denote the fitting error to the atom loss dynamics.
The inset shows the typical observed atom loss dynamics, where the orange solid curve shows the fit result by a double exponential function.
Error bars denote the standard deviation of three scans (smaller than the plots).}
\end{figure}

To induce on-site two-body losses as dissipation, we apply the PA laser which is detuned by -5.76~GHz from the ${^1}S_0\leftrightarrow {^3}P_1\ (F'=7/2)$ transition.
We measure the on-site two-body loss rate due to the PA laser in an isolated double-well lattice of $s_L=(80, 80, (20.8, 20))$, into which evaporatively cooled six-component $^{173}$Yb atoms are loaded.
Here, different from the case of the spin correlation measurement, we load a sufficient number of atoms to prepare doubly occupied sites.
In Fig.~\ref{Fig.S2}, we show the results of the loss rate measurement and a typically observed atom loss dynamics (inset). Here, we fit the data with a double exponential function where the initial faster decay corresponds to the on-site two-body loss and the later slower decay corresponds to the one-body loss mainly due to the photon scattering.
We note that we observe the saturation of loss rate $\gamma_{\text{PA}}$ at a large intensity of the PA laser, which has also been observed in previous works \cite{doi:10.1126/sciadv.1701513,PhysRevA.66.061403,PhysRevA.71.013417,Haimberger_2006,PhysRevLett.101.060406}, and we only use the PA intensities in the unsaturated regime.
The PA light does not play a role in one-body losses or the heating effect since the photon-scattering rate for atoms of single occupation at intensity $\sim 20$~W/cm$^2$ is calculated to be $\gamma_{\text{photon}}=0.021$~/ms, which ensures that the effect of photon scattering is negligible in the time scale that we measure the dynamics in this work ($<$ 2~ms).
Note that we confirm that the PA line for dissipation is active for all spin pairs.
A slight spin-pair dependence of the PA rate observed within the experimental uncertainty is not evident in the STO signals, and therefore, in this work, we analyze the data by assuming no spin-pair dependence and using Eq.~(4).
Here, we note that a different PA-rate measurement in a deep 3D lattice configuration, in which the inter-site hopping is negligible, gives a higher PA rate after the compensation for the difference in the spatial spread of the Wannier function.
This may partially explain the observed discrepancy in the values between the experimental data and theoretical calculation in Fig.~3 of the main text.

\section{S.4 Loading schemes to realize three different initial antiferromagnetic conditions}
\begin{figure}
  \includegraphics[width=8.54cm]{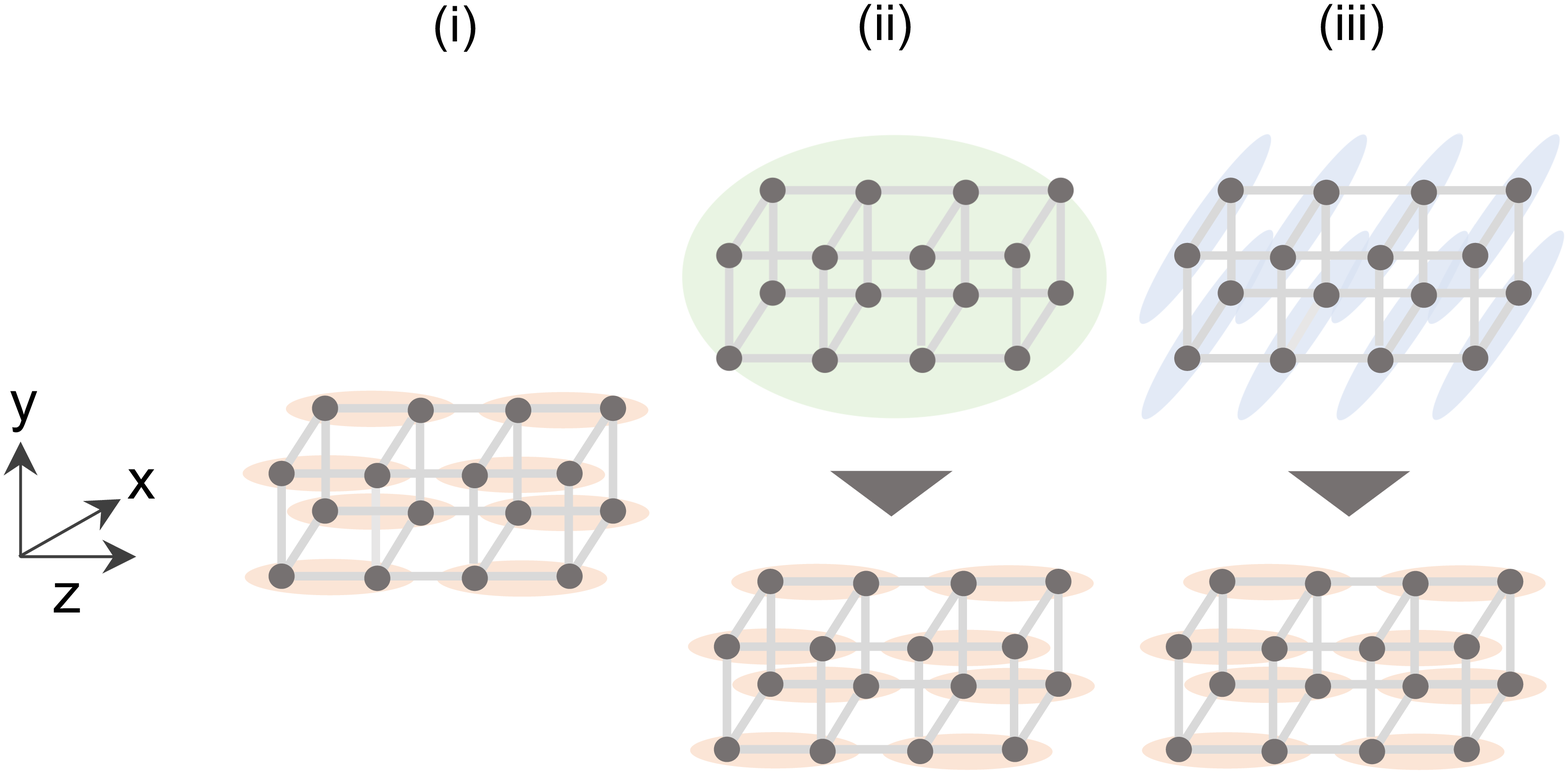}
  \caption{\label{Fig.S3} Schematic illustration of three different atom-loading schemes into isolated double-well lattices: (i) Load to isolated double-well lattices along $z$-axis from the beginning. (ii) First load to a three-dimensional lattice and switch to isolated double-well lattices along $z$-axis in 0.5~ms. (iii) First load to one-dimensional lattices along $x$-axis and switch to isolated double-well lattices along $z$-axis in 0.5~ms.}
\end{figure}

In the investigation of the dynamics in a dissipative isolated double-well lattice, we prepare three systems with different initial antiferromagnetic (AFM) conditions (i), (ii), and (iii) in Fig.~2.
These are realized by different atom-loading schemes to an isolated double-well lattice.
The specific sequence is as follows (Fig.~\ref{Fig.S3}):
\begin{itemize}
  \item[(i)] Large AFM condition (normalized STO amplitude $A\sim 0.3$).\\
  Load to isolated double-well lattices along $z$-axis, which has a lattice depth of $s_L=(80, 80, (20.8, 20))$, from the beginning.

  \item[(ii)] Medium AFM condition ($A\sim 0.2$)\\
  First load to a three-dimensional lattice and switch to isolated double-well lattices along $z$-axis in 0.5~ms : $s_L=(28,  28, (28, 0))\rightarrow(80, 80, (20.8, 20))$.

  \item[(iii)] Small AFM condition ($A\gtrsim 0.0$)\\
  First load to one-dimensional lattices along $x$-axis and switch to isolated double-well lattices along $z$-axis in 0.5~ms : $s_L=(19.8, 80, (80, 0))\rightarrow(80, 80, (20.8, 20))$.
\end{itemize}
In the measurement of spin correlations in the crossover from an isolated double-well lattice to a uniform one-dimensional lattice, we prepare systems with small initial AFM correlations by a loading scheme similar to (iii).

\section{S.5 Spin temperature}
\begin{figure}
  \includegraphics[width=8.54cm]{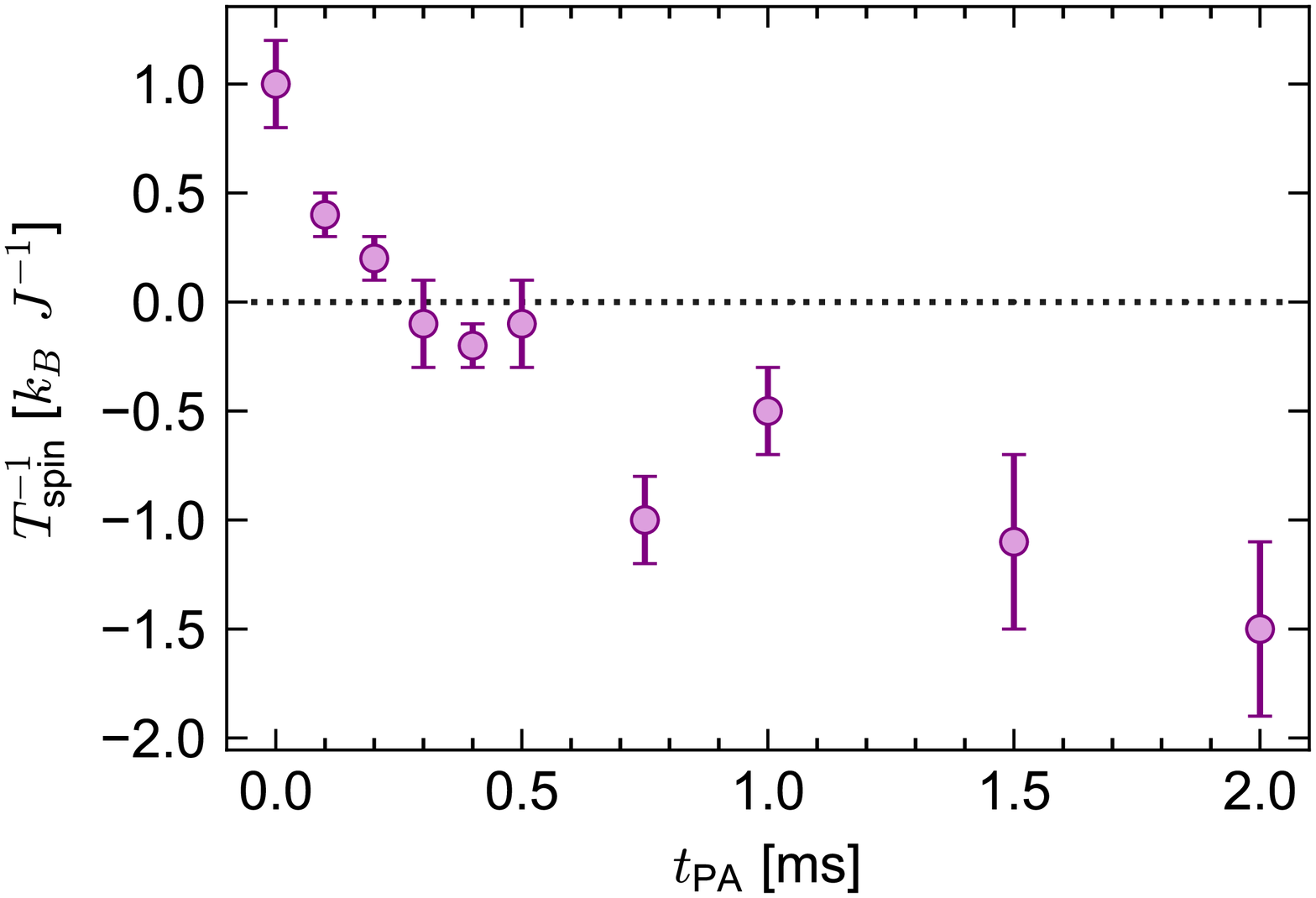}
  \caption{\label{Fig.S4}Dynamics of the inverse of the spin temperature $T_{\text{spin}}$ calculated from the result for the initial AFM condition (ii) shown in Fig.~2(b) and 2(c).
Here, we show the dependency of the inverse of $T_{\text{spin}}$ on the irradiation time $t_{\text{PA}}$ of the PA laser.
Error bars denote the standard deviation, which is calculated from the standard deviations of $p_s$ and $p_{t_0}$.}
\end{figure}

In the state with sign-reversed magnetic correlations, the population of triplet states is greater than that of singlet states, which corresponds to a negative temperature state \cite{PhysRev.103.20}.
Here, we define the spin temperature $T_{\text{spin}}$ as
\begin{equation}
  \frac{p_{t_0}}{p_s}=\exp\left( -\frac{J}{k_B T_{\text{spin}}} \right),
\end{equation}
where $k_B$ is the Boltzmann constant, $p_s/p_{t_0}$ is the singlet/triplet fraction, and $J$ is the exchange interaction.
In Fig.~\ref{Fig.S4}, we show the result of the dynamics of $1/T_{\text{spin}}$ measured in a dissipative isolated double-well lattice, which corresponds to the result for the initial AFM condition (ii) shown in Fig.~2.
As dissipation is introduced, $1/T_{\text{spin}}$ reverses from positive to negative via zero at which $p_s=p_{t_0}$.
This result demonstrates that a negative temperature state, which has been realized in an isolated system \cite{doi:10.1126/science.1227831}, can also be realized in an open quantum system with dissipation.

\section{S.6 Singlet loss rate at finite U/$\lowercase{t_{\bf d}}$}
In the main text (Fig.~3), we investigate the singlet loss rate in an isolated double-well lattice.
To theoretically evaluate the singlet loss rate at finite $U/t_d$, we consider a two-site Fermi-Hubbard model with on-site two-body loss.
The dynamics of the density matrix $\hat{\rho}$ of this system is described by the quantum master equation (Eq.~(1) in the main text).
Here, we use the quantum-trajectory method \cite{doi:10.1080/00018732.2014.933502}, in which the system evolves under the effective non-Hermitian Hamiltonian
\begin{align}
\hat{H}_{\mathrm{eff}}=&\hat{H}-\frac{i\hbar}{4}\sum_{j}\sum_{\sigma \neq \sigma'}\hat{L}_{j\sigma\sigma'}^\dag \hat{L}_{j\sigma\sigma'}\notag\\
=&-t_d\sum_\sigma\left(\hat{c}_{1\sigma}^\dag \hat{c}_{2\sigma}+\mathrm{H.c.}\right)+\frac{\tilde{U}}{2}\sum_{j=1,2}\sum_{\sigma\neq\sigma'}\hat{n}_{j\sigma}\hat{n}_{j\sigma'}
\label{eq_Heff_app}
\end{align}
during a time interval between loss events.
The imaginary part of the complex-valued interaction strength $\tilde{U}:=U-i\hbar\gamma$ is due to the inelastic interaction that causes on-site two-body loss \cite{PhysRevLett.124.147203}.
If the system has two particles with spin $\sigma_1$ and $\sigma_2$ ($\sigma_1\neq\sigma_2$), the eigenvalues $E_n$ and the corresponding right eigenstates $\ket{n}_R$ of the non-Hermitian Hamiltonian \eqref{eq_Heff_app} are given by
\begin{align}
E_0=&\frac{1}{2}(\tilde{U}-\sqrt{\tilde{U}^2+16t_d^2}),\\
\ket{0}_R=&-\frac{2t_d}{E_0}(\ket{\sigma_1,\sigma_2}-\ket{\sigma_2,\sigma_1})+\ket{0,\sigma_1\sigma_2}+\ket{\sigma_1\sigma_2,0},\\
E_1=&0,\\
\ket{1}_R=&\ket{\sigma_1,\sigma_2}+\ket{\sigma_2,\sigma_1},\\
E_2=&\tilde{U},\\
\ket{2}_R=&\ket{0,\sigma_1\sigma_2}-\ket{\sigma_1\sigma_2,0},\\
E_3=&\frac{1}{2}(\tilde{U}+\sqrt{\tilde{U}^2+16t_d^2}),\\
\ket{3}_R=&-\frac{2t_d}{E_3}(\ket{\sigma_1,\sigma_2}-\ket{\sigma_2,\sigma_1})+\ket{0,\sigma_1\sigma_2}+\ket{\sigma_1\sigma_2,0},
\end{align}
where
\begin{align}
\ket{\sigma_1,\sigma_2}=&\hat{c}_{1\sigma_1}^\dag\hat{c}_{2\sigma_2}^\dag\ket{\mathrm{vac}},\notag\\
\ket{\sigma_2,\sigma_1}=&\hat{c}_{1\sigma_2}^\dag\hat{c}_{2\sigma_1}^\dag\ket{\mathrm{vac}},\notag\\
\ket{\sigma_1\sigma_2,0}=&\hat{c}_{1\sigma_1}^\dag\hat{c}_{1\sigma_2}^\dag\ket{\mathrm{vac}},\notag\\
\ket{0,\sigma_1\sigma_2}=&\hat{c}_{2\sigma_1}^\dag\hat{c}_{2\sigma_2}^\dag\ket{\mathrm{vac}},\notag
\end{align}
and $\ket{\mathrm{vac}}$ denotes the vacuum state.
Note that these eigenstates are not normalized.
We also note that the left eigenstates $\ket{n}_L$ are obtained by replacing $\tilde{U}$ in $\ket{n}_R$ by $\tilde{U}^*$, and we thus have $\ket{1}_L=\ket{1}_R$ and $\ket{2}_L=\ket{2}_R$.

Using the non-Hermitian Hamiltonian, we consider the time evolution of the ground state of $\hat{H}$, i.e., the singlet state
\begin{align}
\ket{\psi(0)}=&\frac{1}{\mathcal{N}}\Bigl[\frac{4t_d}{-U+\sqrt{U^2+16t_d^2}}(\ket{\sigma_1,\sigma_2}-\ket{\sigma_2,\sigma_1})\notag\\
&+\ket{0,\sigma_1\sigma_2}+\ket{\sigma_1\sigma_2,0}\Bigr],
\label{eq_initial_state}
\end{align}
where $\mathcal{N}$ is a normalization factor that ensures $\braket{\psi(0)|\psi(0)}=1$.
In the dynamics under $\hat{H}_{\mathrm{eff}}$, the squared norm of a state is not conserved due to non-Hermiticity. The squared norm, which is initially normalized to unity, corresponds to the probability of having no loss events during the dynamics \cite{doi:10.1080/00018732.2014.933502}.
Thus, the decay rate of the squared norm of the singlet state \eqref{eq_initial_state} corresponds to the decay rate of the singlet fraction $p_s$.
Using the completeness relation $\sum_n\ket{n}_{R\ L}\bra{n}/_L\braket{n|n}_R=1$, we can write the time evolution of the squared norm of the solution of the Schr\"{o}dinger equation $i\hbar d\ket{\psi(t)}/dt=\hat{H}_{\mathrm{eff}}\ket{\psi(t)}$ as
\begin{align}
\braket{\psi(t)|\psi(t)}=&\sum_{m,n}c_{m,n}e^{i(E_m^*-E_n)t/\hbar}\notag\\
=&c_{0,0}e^{2\mathrm{Im}[E_0]t/\hbar}+c_{3,3}e^{2\mathrm{Im}[E_3]t/\hbar}\notag\\
&+c_{0,3}e^{i(E_0^*-E_3)t/\hbar}+c_{3,0}e^{i(E_3^*-E_0)t/\hbar},
\label{eq_norm_expansion}
\end{align}
where
\begin{equation}
c_{m,n}:=\frac{\braket{\psi(0)|m}_{L\ L}\braket{n|\psi(0)}}{{}_R\braket{m|m}_{L\ L}\braket{n|n}_R}{}_R\braket{m|n}_R,
\end{equation}
and we have used ${}_L\braket{1|\psi(0)}={}_L\braket{2|\psi(0)}=0$ in the second line of Eq.~\eqref{eq_norm_expansion}.
For the parameters used in our experiment, we find $|c_{3,3}|, |c_{0,3}|, |c_{3,0}| \ll c_{0,0}$ and $c_{0,0}\simeq 1$. Thus, the time evolution of the squared norm is approximated as
\begin{equation}
\braket{\psi(t)|\psi(t)}\simeq e^{-2\Gamma t/\hbar},
\end{equation}
and the singlet decay rate is given by $2\Gamma/\hbar$ with
\begin{align}
\Gamma=&-\mathrm{Im}[E_0]\notag\\
=&\frac{1}{2}\left(\hbar\gamma-\sqrt{\frac{-x+\sqrt{x^2+y^2}}{2}}\right),
\label{eq_Gamma}
\end{align}
where $x:=U^2+16t_d^2-(\hbar\gamma)^2$ and $y:=-2\hbar U\gamma$.

\begin{figure}
  \includegraphics[width=8.54cm]{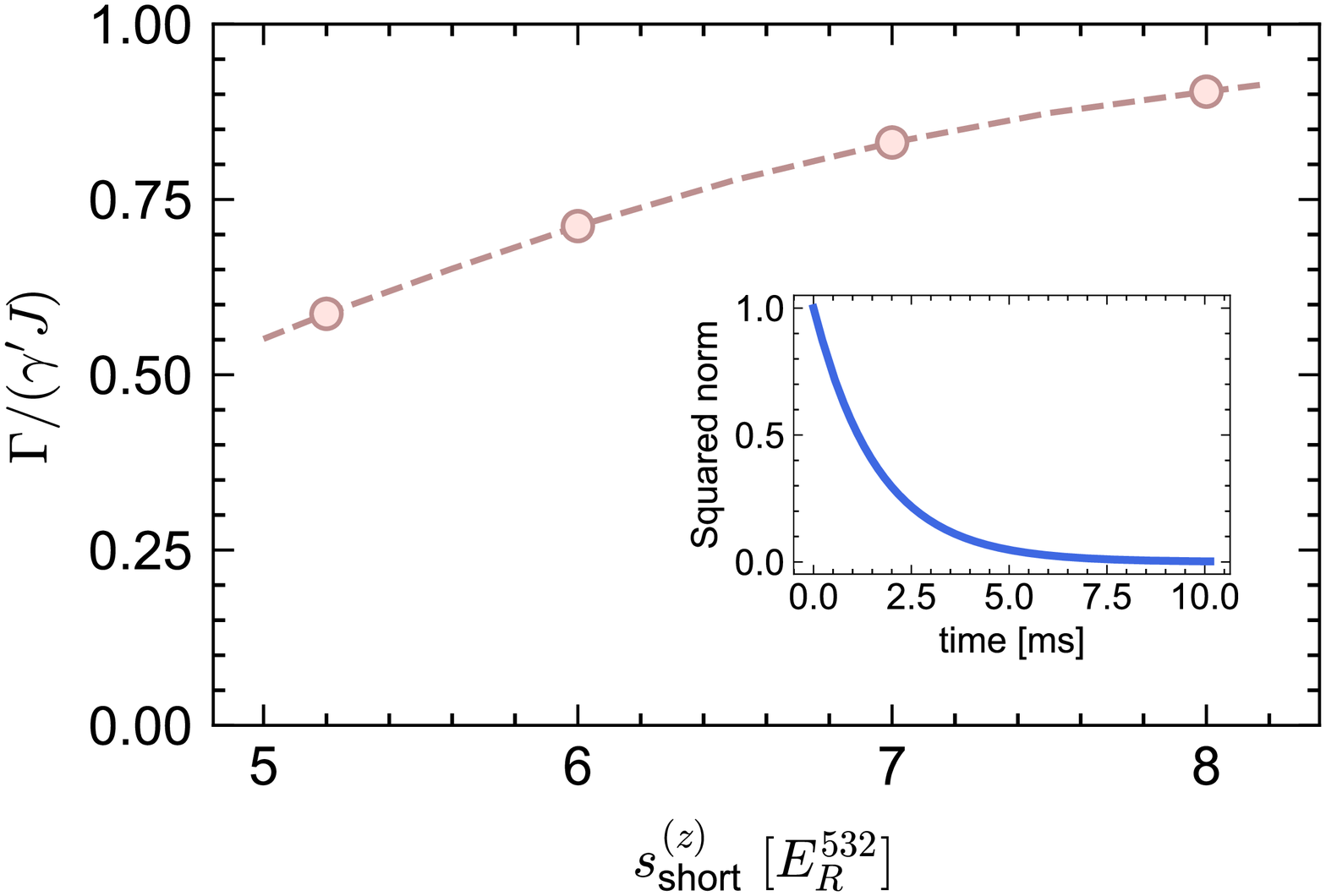}
  \caption{\label{Fig.S5}Dependency of $\Gamma/(\gamma' J)$ on the short-lattice depth.
  Here, $\Gamma$ is half of the singlet decay rate times $\hbar$ calculated with Eq.~\eqref{eq_Gamma}, $\gamma'=\hbar\gamma/U$ is the dimensionless dissipation strength fixed to 0.082(4), and $J=4t_d^2/U$ is the exchange interaction.
  The dots correspond to the values for $s^{(z)}_{\mathrm{short}}~=~5.2, 6.0, 7.0, 8.0$.
  The inset shows a typical result for the dynamics of the squared norm of the singlet state \eqref{eq_initial_state} under $\hat{H}_{\mathrm{eff}}$ calculated with exact diagonalization for the parameters $U/t_d = 4.0$ and $\gamma'=0.082$.}
\end{figure}

In the main text, we find that the decay rate of $p_s$ increases approximately in proportion to the exchange interaction $J=4t_d^2/U$.
Such a linear dependence is expected for $U\gg t_d$, whereas $U/t_d$ takes a value between 4.0 and 10.9 in our case. We confirm that $\Gamma$ is approximately proportional to $J$ under the fixed dimensionless dissipation strength $\gamma'=\hbar\gamma/U$ even in the regime of $U\gtrsim t_d$.
In Fig.~\ref{Fig.S5}, we show the dependence of
\begin{widetext}
\begin{align}
\frac{\Gamma}{\gamma'J}=&\frac{1}{8}\Bigl(\frac{U}{t_d}\Bigr)^2-\frac{1}{8\gamma'}\Bigl(\frac{U}{t_d}\Bigr)^2\sqrt{\frac{-1-16(t_d/U)^2+(\gamma')^2+\sqrt{(1+16(t_d/U)^2-(\gamma')^2)^2 + 4(\gamma')^2}}{2}}
\end{align}
\end{widetext}
on the short-lattice depth $s^{(z)}_{\mathrm{short}}$.
The dependency of $\Gamma/(\gamma' J)$ on $s^{(z)}_{\mathrm{short}}$ is small (deviation/mean $<$ 21\%), meaning that the plots of the singlet decay rate $p_s$ against $J$ fall approximately on the same line.


\providecommand{\noopsort}[1]{}\providecommand{\singleletter}[1]{#1}%
%